\documentstyle[12pt,aaspp4]{article}

\def\noteforeditor#1{{}}
\def\firstuse#1{}
\slugcomment{To appear in the August 2000 issue of the {\em PASP}}

\lefthead{MIGHELL}
\righthead{ASTROMETRY OF THE $\omega$ CEN HST CALIBRATION FIELD}

\begin{document}

\def\et{\hbox{et~al.\ }}
\def\vs{\hbox{vs }}
\def\feh{\hbox{[Fe/H]}}
\def\bmv{\hbox{$B\!-\!V$}}
\def\vmi{\hbox{$V\!-\!I$}}
\def\ebmv{\hbox{$E(\bmv)$}}
\def\dbmv{\hbox{$d_{(B\!-\!V)}$}}
\def\dbmr{\hbox{$d_{(B\!-\!R)}$}}
\def\mvto{\hbox{$M_V^{\rm{TO}}$}}
\def\mvrhb{\hbox{$M_V^{\rm{RHB}}$}}
\def\vhb{\hbox{$V_{\rm{HB}}$}}
\def\lea{\mathrel{<\kern-1.0em\lower0.9ex\hbox{$\sim$}}}
\def\gea{\mathrel{>\kern-1.0em\lower0.9ex\hbox{$\sim$}}}
\def\ifundefined#1{\expandafter\ifx\csname#1\endcsname\relax}

\title{Astrometry of the $\omega$ Centauri {\sl{Hubble Space Telescope}}\\
Calibration Field\altaffilmark{1}
}

\author{
{\sc
Kenneth J. Mighell\altaffilmark{2}
}
}
\affil{Kitt Peak National Observatory\\
National Optical Astronomy Observatories\altaffilmark{3}\\
P.~O.~Box 26732, Tucson, AZ~~85726\\
{\em mighell@noao.edu}
}

\altaffiltext{1}{
Based on observations made with the NASA/ESA
{\sl{Hubble Space Telescope}},
obtained from the data archive at the Space Telescope
Science Institute,
which is operated by the Association of
Universities for Research in Astronomy, Inc.\ under NASA
contract NAS5-26555.
}

\altaffiltext{2}{
Guest User, Canadian Astronomy Data Centre, which is operated by the
Dominion Astrophysical Observatory for the National Research Council of
Canada's Herzberg Institute of Astrophysics.
}

\altaffiltext{5}{NOAO is operated
by the Association of Universities for Research in Astronomy, Inc., under
cooperative agreement with the National Science Foundation.}

\begin{abstract}
Astrometry,
on the International Celestial Reference Frame (epoch J2000.0),
is presented for the Walker (1994, \pasp, 106, 828) stars
in the $\omega$ Centauri
(=NGC\,5139=C\,1323-1472)
{\sl{Hubble Space Telescope}}
Wide Field/Planetary Camera (WF/PC)
calibration field
of Harris \et (1993, \aj, 105, 1196).
Harris \et stars were first identified on a
WFPC2 observation of the $\omega$ Cen {\sl HST} calibration field.
Relative astrometry of the Walker stars in this field
was then obtained using Walker's
CCD positions and astrometry derived using the {\sc{stsdas}}
{\sc{metric}} task on the positions of the Harris \et stars
on the WFPC2 observation.
Finally, the relative astrometry,
which was based on the {\sl{HST}} Guide Star Catalog,
is placed on the International Celestial Reference Frame
with astrometry from the USNO-A2.0 catalog.
An ASCII text version of the
astrometric data of the Walker stars in the $\omega$ Cen {\sl HST}
calibration field is available
electronically in the online version of the article.
\end{abstract}


\clearpage
\section{INTRODUCTION}

Walker (\cite{W94}; hereafter W94)
presented ground-based $BV$ (Johnson) $RI$ (Kron-Cousins) photometry
of the Harris \et (\cite{H93}; hereafter H93)
{\sl{Hubble Space Telescope (HST)}}
Wide-Field/Planetary Camera (WF/PC) calibration field
in the Galactic globular cluster $\omega$ Centauri
(=NGC\,5139=C\,1323-1472).
This small (100\arcsec$\!\times$100\arcsec)
field is located 13.2 arcmin southwest of the center of $\omega$ Cen at
13$^{\rm{h}}$\,25$^{\rm{m}}$\,37$^{\rm{s}}$
-47\arcdeg\,35\arcmin\,38\arcsec\ (epoch J2000.0)
[see Fig.\ 1 (Plate 59) of H93, Figs.\ 1 and 2 of W94].
Although this field has been observed over 1100 times
with the {\sl{HST}}, there is no astrometry available for the W94
stars in the published literature.
In an effort to encourage various ongoing WFPC2 calibration efforts, I now
present astrometry (J2000.0)
of the W94 stars in the $\omega$ Cen {\sl HST} calibration field
on the International Celestial Reference Frame.

\section{THE HARRIS ET AL. (1993) STARS}

The first step of the
astrometric calibration of the W94 stars in the $\omega$ Cen
{\sl HST} calibration field
is to identify which H93 stars given in Table 4 of W94 are present
on a WFPC2 observation of that field.

Table \ref{tbl-01}\firstuse{Table\ref{tbl-01}}\ is the combination of
all the stars in Table 4 of W94 with $x$ and $y$ CCD positions
along with their astrometry as given in Table 1 of H93.
Columns 1--7 of Table \ref{tbl-01} are from Table 4 of W94;
columns 8 and 9 give the right ascensions and declinations (J2000.0)
as listed in Table 1 of H93; column 10 is the identification number
given in Table 5 of W94.

I determined a plate solution for the W94 observations
using the data in Table \ref{tbl-01} and
the {\sc{iraf}} {\sc{ccmap}} task.
A simple linear model in $x$ and $y$ was fitted
(6 unknowns) to a tangent sky projection geometry.
The {\sc{ccmap}}
fit results are presented in the first section of
Table \ref{tbl-02}\firstuse{Table\ref{tbl-02}}.
Column 1 of Table \ref{tbl-02}
gives information about the input positions and astrometry;
column 2 gives the number of stars in the input dataset;
columns 3 and 4 gives the fit solution for the plate scale in
units of arcsec per pixel (arcsec px$^{-1}$);
columns 5 and 6 gives the fit solutions for axis rotation in degrees;
columns 7 and 8 gives the fit rms for the standard astrometric coordinates
$\xi$ and $\eta$ in units of arcsec.

Walker states that the size of his pixels was 0.38 arcsec.
This agrees well with the fitted plate scale of 0.39 arcsec px$^{-1}$
given in the first section of Table \ref{tbl-02}.
We see
in Fig.\ 1 of W94
that the orientation of the W94 data is
North to the left (low values of $x$) with East at the bottom
(low values of $y$).
This agrees well with the fitted axis rotations of
$\sim$88\arcdeg\ and $\sim$268\arcdeg\ given in the
first section of Table \ref{tbl-02}.
This suggests that the CCD camera used by Walker was just slightly out of
alignment from a perfect East-West orientation which
would have given values of 90\arcdeg\ and 270\arcdeg, respectively,
for axis rotation.
The respective fit rms values
for the standard astrometric coordinates $\xi$ and $\eta$ was
only 73.0 and 40.7 milliarcsec (mas) for the
first fit with all 22 stars in Table \ref{tbl-01};
a smaller subset of 12 stars improves the respective rms fits values to
21.3 and 33.0 mas.

I chose two WFPC2 exposures,
{\tt{U4PH0101R.C0H}} (14-s F814W)
and
{\tt{U4PH0102R.C0H}} (100-s F814W),
which observed the $\omega$ Cen {\sl HST} calibration
field with the WF2 aperture
(Biretta \et \cite{biretta_et_al_1996}) at the same position and
same roll angle.
These observations were secured as
part of the {\sl HST} Cycle 7 WFPC2 calibration
program CAL/WF2 7929 (PI: Casertano)
and were placed in the public data archive
at the Space Telescope Science Institute on 1998 March 31.
The datasets were recalibrated at the Canadian Astronomy Data Centre
and retrieved electronically using a guest account
which was established for this purpose.

The astrometric calibration of the W94 stars
began by identifying which of the
H93 stars in Table \ref{tbl-01} were present on the WF2 section of
WFPC2 observation {\tt{U4PH0101R.C0H}}.
This was accomplished
by applying the previously derived plate solutions
to Table \ref{tbl-01} using
the {\sc{iraf}} {\sc{cctran}} task.
Even the first fit with all 22 H93 stars produced a
plate solution which allowed for the unambiguous identification of
the H93 stars present on the WF2 section
of this 14-s WFPC2 observation.
Some of the H93 stars in Table \ref{tbl-01}
(e.g. H93 star 917 = W94 star 30)
do not appear on the WF2 section of this WFPC2
observation; some are probable ground-based blends
(e.g. H93 star 1368 = W94 star 23);
some have WFPC2 stellar images are distorted by WFPC2 CCD defects
(e.g. H93 star 1200 = W94 star 33);
and some are so bright that they show diffraction spikes even on a
short 14-s exposure
(e.g. H93 star 1629 = W94 star 1).

The following list of 12 H93 stars is a clean subset of Table 1
which produced the best plate solution of the first section of
Table \ref{tbl-02} with rms fitting errors
of $\sigma_\xi=21.3$ mas (0.055 px)
and $\sigma_\eta=33.0$ mas (0.085 px) :
903,
1035,
1056,
1063,
1249,
1461,
1462,
1512,
1522,
1565,
1655,
1686.
This fit had a plate scale which was identical in both directions
(0.388 arcsec px$^{-1}$) with axis rotation values of
88.056\arcdeg\ and 268.079\arcdeg.

\section{RELATIVE ASTROMETRY OF THE WALKER (1994) STARS}

The next step of the astrometric calibration of the W94 stars
in the $\omega$ Cen {\sl HST} calibration field
is to obtain relative astrometry
of all the stars in Table 5 of W94.

Table \ref{tbl-03}\firstuse{Table\ref{tbl-03}}\ gives
the W94 $x$ and $y$ CCD positions
and the relative astrometry of the clean subset of 12 H93
stars identified in the previous section.
Columns 1--4 of Table \ref{tbl-03} are from Table \ref{tbl-01};
columns 7 and 8 give the $x$ and $y$ positions of these stars on the WF2
section of the WFPC2 observation {\tt{U4PH0101R.C0H}} as determined using
the {\sc{iraf}} {\sc{imexamine}} task;
columns 5 and 6 give the right ascensions and declinations (J2000.0)
as determined with the {\sc{stsdas}} {\sc{metric}} task using the
{\tt{U4PH0101R.C0H}} observation with
the WF2 CCD positions given in columns 7 and 8.

I determined a plate solution for the W94 observations
using {\sc{ccmap}} with the data in Table \ref{tbl-03}.
A simple linear model in $x$ and $y$ was fitted
(6 unknowns) to a tangent sky projection geometry.
The {\sc{ccmap}}
fit results are presented in the second section of Table \ref{tbl-02}.
This plate solution had rms fitting errors
of $\sigma_\xi=20.3$ mas (0.052 px)
and $\sigma_\eta=22.3$ mas (0.058 px);
fitted plate scales of 0.387 arcsec px$^{-1}$ for both $x$ and $y$;
and axis rotation values of 88.536\arcdeg\ and 268.583\arcdeg.
Note that this result is nearly identical to that of the
best plate solution derived in the previous section.

I determined relative astrometry (J2000.0) for all of the W94 stars
by using {\sc{cctran}} with this plate solution and the
$x$ and $y$ CCD positions given in Table 5 of W94.
Figure \ref{fig-01}\firstuse{Fig\ref{fig-01}}\ shows
the W94 stars ({\em{circles}})
which are present on the WF2 section of the 100-s WFPC2
{\tt{U4PH0102R.C0H}} observation; the plotted positions of these W94 stars
was determined with the {\sc{stsdas}} {\sc{invmetric}} task.
The two WFPC2 observations,
{\tt{U4PH0101R.C0H}} and {\tt{U4PH0102R.C0H}},
have identical astrometry; I used the deeper observation in Fig.\ \ref{fig-01}
in order to better display the location of the faintest W94 stars.

\section{ASTROMETRY OF THE WALKER (1994) STARS ON THE ICRF}

The final step of the astrometric calibration of the W94 stars
in the $\omega$ Cen {\sl HST} calibration field
is to place the relative astrometry
of the W94 stars on the International Celestial Reference Frame
(ICRF)\footnote{The {\sl{Hipparcos}} catalog to the non-specialist.}.

The plate solution derived in the previous section
is ultimately based on
the {\sl{HST}} Guide Star Catalog which is known to have stars
with large positional errors.
I demonstrate below that the plate solution determined in the previous section
systematically deviates from the ICRF (J2000.0) by about 1.1 arcsec.

I used the USNO-A2.0 finder chart
website\footnote{\tt{http://ftp.nofs.navy.mil/data/fchpix/}}
to find the 97 USNO-A2.0 (Monet et al.\ \cite{monet_et_al_1998})
astrometric catalog stars within a
2\arcmin$\times$2\arcmin\ box centered on the field center of the
$\omega$ Cen {\sl HST} calibration field
(see Table \ref{tbl-04}\firstuse{Table\ref{tbl-04}}).
Columns 2 and 3 of Table \ref{tbl-04}
give the right ascensions and declinations (J2000.0)
on the International Celestial Reference Frame (ICRF);
column 4 gives the red plate magnitude; column 1 gives my identification
number for these nearby USNO-A2.0 stars.

Table \ref{tbl-05}\firstuse{Table\ref{tbl-05}}\ combines the
W94 $x$ and $y$ CCD positions
with astrometry from the USNO-A2.0 catalog for 20 W94 stars
with probable USNO-A2.0 counterparts.
Columns 1--3 of Table \ref{tbl-05} are from Table 5 of W94;
column 6 gives the my identification number of the USNO-A2.0 star in
Table \ref{tbl-04} which I associate with the W94 star;
columns 4--5 give the right ascensions and declinations (J2000.0)
given in Table \ref{tbl-04}.

I determined plate solutions for the W94 observations
using the data in Table \ref{tbl-05} and
the {\sc{ccmap}} task.
A simple linear model in $x$ and $y$ was fitted
(6 unknowns) to a tangent sky projection geometry.
The {\sc{ccmap}}
fit results are presented in the third section of
Table \ref{tbl-02}.
The plate solution with 20 W94 stars had rms fitting errors
of $\sigma_\xi=280$ mas (0.72 px)
and $\sigma_\eta=467$ mas (1.2 px).
The fitting error for $\eta$ is considerably larger than the nominal
250 mas positional error
for the USNO-A2.0 catalog.
The plate solution based on a unambiguous
subset of 12 W94 stars had rms fitting errors
of $\sigma_\xi=225$ mas (0.58 px)
and $\sigma_\eta=248$ mas (0.64 px)
which does agree with the
nominal USNO-A2.0 positional errors.

The plate solution based on Table \ref{tbl-03} predicts a position
for W94 star 2 (= H93 star 1438) of
13$^{\rm{h}}$\,25$^{\rm{m}}$\,35\,\fs869
-47\arcdeg\,35\arcmin\,27\,\farcs66 (J2000.0).
I associate this star with the 38th star in
Table \ref{tbl-04} which has the USNO-A2.0 catalog position of
13$^{\rm{h}}$\,25$^{\rm{m}}$\,35\,\fs9327
-47\arcdeg\,35\arcmin\,28\,\farcs6580 (J2000.0).
The difference between these positions in
right ascension and declination are
$\Delta \alpha_{2000} = 0\,\fs0637$
and
$\Delta \delta_{2000} = -0\,\farcs92$, respectfully,
giving an angular separation of $\sim$1.12 arcsec.

Observing that the USNO-A2.0 stars in Fig.\ 1 are systematically
offset from the stellar images by $\sim$1.1 arcsec, we come to the
odd conclusion that
Table \ref{tbl-02} indicates that the plate solutions based on
H93 astrometry (Table \ref{tbl-01}) and
{\sl{HST}} {\sc{metric}} astrometry (Table \ref{tbl-03})
are considerably more precise
(i.e., have excellent fitting errors)
--- but are significantly less accurate
(i.e., have large systematic offsets)
--- than the plate solutions based
on USNO-A2.0 astrometry (Table \ref{tbl-05}).

The precise relative astrometry derived in the previous section
can be placed on the International Celestial Reference Frame
by simply adding 0\,\fs0637 and -0\,\farcs92
to the right ascensions and declinations derived from the plate
solution based on {\sc{metric}} astrometry of the WF2 section
of {\sl{HST}} WFPC2 observation {\tt{U4PH0101R.C0H}}
(see Table \ref{tbl-06}\firstuse{Table\ref{tbl-06}}).
The squares shown on Fig.\ \ref{fig-01} indicate the ICRF positions
of the W94 stars; these are offset from the stellar images by $\sim$1.1
arcsec {\em{on this particular image}}.
The absolute rms errors of the J2000.0 right ascensions and declinations
given in Table \ref{tbl-06} are estimated to be
0.35 angular arcsec.
This estimate is
the sum of the nominal 0.25 arcsec positional
error of the USNO-A2.0 catalog and the estimated relative
rms positional error of approximately 0.10 arcsec due to the
uncertainty of the WFPC2 astrometric calibration as a function of time.

Other WFPC2 observations of this field obtained with different
guide stars, roll angles, and apertures,
may exhibit different offsets between the stellar images on a given WFPC2 CCD
and the computed $(x,y)$ pixel locations of the W94 stars on that chip
as derived using the {\sc{invmetric}} task with Table \ref{tbl-06}.

In order to identify the W94 stars on any given WFPC2
observation of the $\omega$ Cen {\sl HST} calibration field,
the reader should first find a given W94 star
(preferably W94 star 2) somewhere in observation and then measure
its $(x,y)$ centroid on the particular chip where the star is
present\footnote{For example, the center of W94 star 2 is
found at the $(x,y)$ location of $(484.92,266.90)$ of
the WF2 section of the WFPC2 observation {\tt{U4PH0102R.C0H}}\,.
}.
Convert that CCD position to a right ascension and declination
using the {\sc{metric}}
task\footnote{The following {\sc{iraf}} command,
{\tt{metric "u4ph0101r.c0h[2]" x=484.92 y=266.90}}~,
gives
$\alpha_{2000} = $13$^{\rm{h}}$\,25$^{\rm{m}}$\,35\,\fs8656
and
$\delta_{2000} = $-47\arcdeg\,35\arcmin\,27\,\farcs792~.
}.
Next, determine the difference in right ascension ($\Delta \alpha_{2000}$)
and declination ($\Delta \delta_{2000}$) between the {\sc{metric}} value
and the ICRF position of that star as given in
Table \ref{tbl-06}\footnote{Using {\sc{iraf}} sexagesimal notation,
we find that
$
\Delta \alpha_{2000}
=
\alpha_{\mbox{\sc{metric}}}
-
\alpha_{\mbox{\sc{table6}}}
=
(\mbox{13:25:35.8656})
-
(\mbox{13:25:35.933})
=
-0\,\fs067
$
and
$
\Delta \delta_{2000}
=
\delta_{\mbox{\sc{metric}}}
-
\delta_{\mbox{\sc{table6}}}
=
(\mbox{-47:35:27.792})
-
(\mbox{-47:35:28.58})
=
+0\,\farcs788$~.
}.
Add $\Delta \alpha_{2000}$ and $\Delta \delta_{2000}\,$
to all the right ascensions and declinations in
Table \ref{tbl-06}\footnote{Continuing with the example,
$\alpha_{\mbox{\sc{table6}}} + \Delta\alpha_{2000} =
\mbox{13$^{\rm{h}}$\,25$^{\rm{m}}$\,35\,\fs866}$
and
$\delta_{\mbox{\sc{table6}}} + \Delta\delta_{2000} =
\mbox{-47\arcdeg\,35\arcmin\,27\,\farcs79}$~.
}.
Finally, determine the $(x,y)$ locations
of the W94 stars on the individual WFPC2 CCDs
using the {\sc{invmetric}}
task with the modified Table \ref{tbl-06} positions\footnote{
The following {\sc{iraf}} command,
{\tt{invmetric u4ph0102r.c0h[2] ra="13:25:35.866" dec="-47:35:27.79"}}~,
gives the $(x,y)$ CCD location of $(484.95,266.93)$ on the WF2 section
of {\tt{U4PH0102R.C0H}}.  The small difference (0.058 px)
between the computed pixel location and the measured value
of $(484.92,266.90)$ is due to (1) centroiding errors,
(2) round-off errors,
and
(3) the small difference between the {\sc{metric}} value,
13$^{\rm{h}}$\,25$^{\rm{m}}$\,35\,\fs8656
-47\arcdeg\,35\arcmin\,27\,\farcs792
(determined above),
and the plate solution value for W94 star 2,
13$^{\rm{h}}$\,25$^{\rm{m}}$\,35\,\fs869
\mbox{-47}\arcdeg\,35\arcmin\,27\,\farcs66~,
which was derived with Table \ref{tbl-03} in the previous section.
}
on a chip-by-chip basis.

\acknowledgments

I am grateful to Alistair Walker for sending me a glossy reprint of W94
as well as electronic versions of Tables 4 and 5 from W94.
I would also like to thank
Lindsey Davis,
Mike Fitzpatrick,
Stephen Levine,
Hugh Harris,
Stefano Casertano,
Brad Whitmore,
J.\,C.~Hsu,
Andy Fruchter
for
discussions
and
advice on various aspects of this project.
This research was supported by a grant from
the National Aeronautics and Space Administration (NASA),
Order No.\ S-67046-F, which was awarded by
the Long-Term Space Astrophysics Program (NRA 95-OSS-16).
This research has made use of
NASA's Astrophysics Data System Abstract Service.

\def\fig01cap{
\label{fig-01}
\noteforeditor{Print this figure TWO (2) COLUMNS wide.\newline}
Part of the $\omega$ Centauri {\sl{Hubble Space Telescope}}
calibration field centered at
13$^{\rm{h}}$\,25$^{\rm{m}}$\,37$^{\rm{s}}$
-47\arcdeg\,35\arcmin\,38\arcsec\ (epoch J2000.0).
This image is the
WF2 section of the 100-s F814W {\sl{HST}} WFPC2 observation
{\tt{U4PH0102R.C0H}}.
The numbered {\em{diamonds}} indicate the positions of the USNO-A2.0 catalog
stars listed in Table 4.
The numbered {\em{circles}} indicate the positions of the stars
in Table 5 of Walker (1994).
Asterisks indicate stars which were used to determine plate solutions
(see Tables \protect\ref{tbl-01}, \protect\ref{tbl-03}, \protect\ref{tbl-05}).
The {\em{squares}} indicate the positions of the stars in Table 5
of Walker (1994) on the International Celestial Reference Frame.
The stellar images are separated from the squares by approximately
1.1 arcsec;
the astrometric system of this particular {\sl{HST}} observation
is systematically displaced from the ICRF by this small angular offset.
The small tickmarks along the sides of the image have a separation
of 100 pixels which is equivalent to an angular separation of 10 arcsec
in the central region of the image.
}
\ifundefined{showfigs}{
  \newpage
  \centerline{{\Large\bf{Figure Captions}}}
  \smallskip
  \figcaption[]{\fig01cap}
}\else{
  \begin{figure}[p]
    \figurenum{1}
    \vspace*{-10mm}
    \hspace*{-8mm}
    \epsfxsize=7.0truein
    \epsfbox{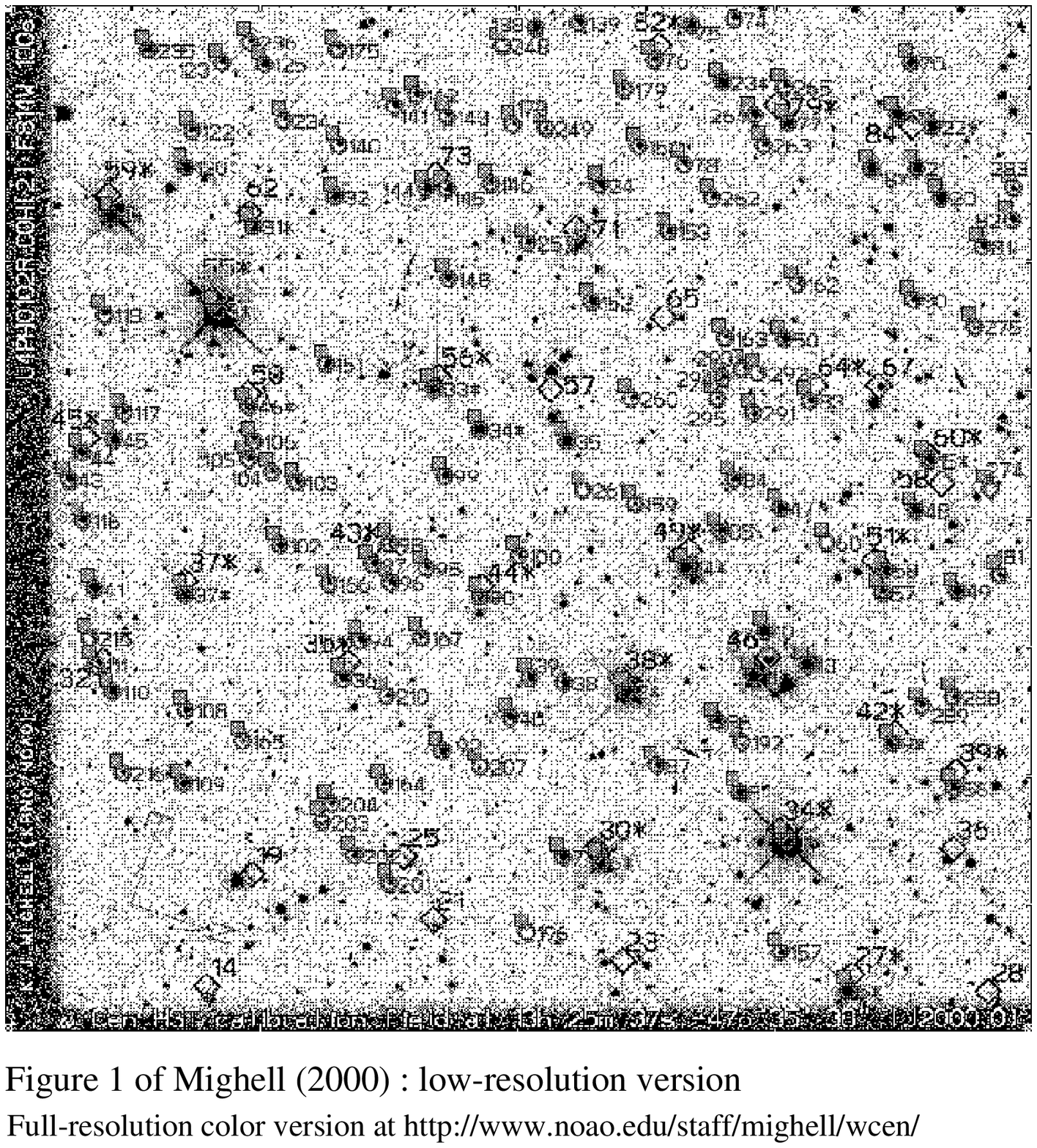}
    \vspace*{-6mm}
    \caption[]{\baselineskip 1.15em \fig01cap}
  \end{figure}
}
\fi

\newpage
\begin{table}
\dummytable\label{tbl-01}
\ifundefined{showtables}{
}\else{
  \vspace*{-30mm}
  \hspace*{-20mm}
  \epsfxsize=8.0truein
  \epsfbox{mighell.tab01.eps}
}
\fi
\end{table}

\newpage
\begin{table}
\dummytable\label{tbl-02}
\ifundefined{showtables}{
}\else{
  \vspace*{-30truemm}
  \hspace*{-20truemm}
  \epsfxsize=8.0truein
  \epsfbox{mighell.tab02.eps}
}
\fi
\end{table}

\newpage
\begin{table}
\dummytable\label{tbl-03}
\ifundefined{showtables}{
}\else{
  \vspace*{-30truemm}
  \hspace*{-20truemm}
  \epsfxsize=8.0truein
  \epsfbox{mighell.tab03.eps}
}
\fi
\end{table}

\newpage
\begin{table}
\dummytable\label{tbl-04}
\ifundefined{showtables}{
}\else{
  \vspace*{-30truemm}
  \hspace*{-20truemm}
  \epsfxsize=8.0truein
  \epsfbox{mighell.tab04_preprint.eps}
}
\fi
\end{table}

\newpage
\begin{table}
\dummytable\label{tbl-05}
\ifundefined{showtables}{
}\else{
  \vspace*{-30truemm}
  \hspace*{-20truemm}
  \epsfxsize=8.0truein
  \epsfbox{mighell.tab05.eps}
}
\fi
\end{table}

\newpage
\begin{table}
\dummytable\label{tbl-06}
\ifundefined{showtables}{
}\else{
  \vspace*{-30truemm}
  \hspace*{-20truemm}
  \epsfxsize=8.0truein
  \epsfbox{mighell.tab06_preprint.eps}
}
\fi
\end{table}

\begin{table}
\dummytable\label{tbl-x}
\end{table}

\end{document}